\documentclass[aps, pra, twocolumn, notitlepage,superscriptaddress,nofootinbib   
]{revtex4-2}
\usepackage{amssymb,latexsym,amsmath}
\usepackage{bm}
\usepackage{amssymb,amsmath}
\usepackage{tensor}
\usepackage{graphicx,xcolor}
\usepackage{float}
\usepackage[titletoc, title]{appendix}
\usepackage[caption=false,lofdepth,lotdepth]{subfig}
\usepackage{soul}
\usepackage[normalem]{ulem}
\usepackage{hyperref}   
\hypersetup{%
	pdfpagemode=FullScreen,
	pdfstartpage=1,
	pdfmenubar=true,
	pdftoolbar=true,
	colorlinks = true,
	linkcolor=blue,
	citecolor=blue,
	urlcolor=blue,
	bookmarksopen=false
}

\allowdisplaybreaks

\begin{document}

\title{Rotational states of an asymmetric vortex pair with mass imbalance in binary condensates}
\author{Alice Bellettini}
\email{alice.bellettini@polito.it}
\affiliation{Department of Applied Science and Technology, Politecnico di Torino, 10129 Torino, Italy}

\author{Andrea Richaud}
\email{andrea.richaud@upc.edu}
\affiliation{Scuola Internazionale Superiore di Studi Avanzati (SISSA), Via Bonomea 265, I-34136, Trieste, Italy}
\affiliation{Departament de F\'isica, Universitat Polit\`ecnica de Catalunya, Campus Nord B4-B5, E-08034 Barcelona, Spain}

\author{Vittorio Penna}
\affiliation{Department of Applied Science and Technology, Politecnico di Torino, 10129 Torino, Italy}

\date{\today}

\begin{abstract}
We consider massive vortices in binary condensates, where the immiscibility condition entails the trapping of the minority component in the vortex cores
of the majority component.
We study such vortices by means of a 2D point-like model, and show how the relevant dynamical equations
exhibit vortex-pair solutions characterized by different vortex masses and circular orbits of different radii $a$ and $b$. 
These solutions are validated by the simulations of the Gross-Pitaevskii equations for binary condensates.
After examining the properties of vortex-pair rotational frequency $\Omega$ as a function of the vortex masses for a given pair geometry, we define the rotational-state diagram $\cal D$, describing all the possible vortex-pair solutions in terms of the orbit radii at given $\Omega$. This includes solutions with equal-mass pairs but $a \ne b$ or with one of the two masses (or both) equal to zero. Also, we analytically find the minimum value of $\Omega$ for the existence of such solutions, and obtain
numerically the critical frequency $\Omega_c$ 
below which $\cal D$ changes its structure and the transition to an unstable
vortex-pair regime takes place.
Our work highlights an indirect measurement scheme to infer the vortex masses from orbits radii $a$ and $b$, and a link between the vortex masses and the vortex-pair small-oscillation properties.
\end{abstract}

\maketitle

\section{Introduction}

In contrast to superfluid helium, Bose-Einstein condensates (BECs) offer a controllable platform for studying superfluidity and quantized vortices \cite{Onsager1949, Fiszdon1991}, the latter being a signature of superfluidity. A review on the vortex states and dynamics in Bose-Einstein condensates can be found in Ref. \cite{Fetter2001}.
Thanks to the much larger length scale characterizing BECs vortices with respect to ${\-}^4$He vortices, the condensates provide a good system for validating theoretical predictions
about vortex phenomenology. 
The single-vortex formation and lattice realization have been first experimentally realized by \cite{Matthews1999, Madison2000, Ketterle2001, Ketterle2001b}. In the case of binary mixtures,
vortices with in-filled core have been experimentally observed \textit{in situ}, via phase contrast techniques \cite{Matthews1999, Anderson2000}. In this case, the vortex cores of the first species are enlarged due to the interaction with the second species.

Multi-component 2D systems offer in fact a varied scenario, where the tuning of the inter-species and intra-species interactions, and the possible manipulation of the hyperfine spin states of atomic species give rise to a rich interplay.
In binary condensates, the minority species \textit{B} sits within the vortex-cores of the majority species \textit{A}. The stability of such ``massive vortices" was studied in Refs. \cite{McGee2001, Law2010}, and \cite{Mukherjee2020}. Besides, in Ref. \cite{Mukherjee2020} the authors characterized the spontaneous formation of different vortex-bright-soliton structures. Additionally, they studied the dynamics of such structures via the analysis of the compressible and incompressible kinetic energy, together with the angular momentum exchange between the two species. 
In the miscible regime, a critical vortex filling for the second component to remain bound within the vortex core was predicted \cite{Gallemí2018}, while in Ref. \cite{Kevrekidis2012} the dynamical behavior of in-phase and out-of-phase vortex-bright-soliton dipoles was studied. 

The interest in the properties of massive vortices has also extended to systems of vortex  necklaces \cite{Caldara2023,Chaika2023} and vortex lattices. Here, triangular, rectangular and square lattices have been studied for miscible \cite{Mueller2002} or attractive mixtures \cite{Kuopanportti2012}, and square vortex lattices in binary BECs were experimentally observed \cite{Schweikhard2004}.
Conversely, in the strongly immiscible regime, the two-component separation was shown \cite{Kasamatsu2009} to lead to vortex sheets within the single-component domains, while
the study of ``dark-bright" solitons in two-component systems via an adiabatic invariant approach \cite{Kevrekidis2018} revealed a behavior analogous to the snake-instability predicted in Ref. \cite{Kuznetsov1995}.

Under another perspective, in the small-mass limit, vortex masses can be viewed as impurities within the vortex cores. Vortex lattices with embedded impurities were explored in Refs. \cite{Caracanhas2015, Characanas2018}.
To complete this rich scenario, it is worth recalling also the spinor-BEC vortices, which were related to topological excitations such as skyrmions \cite{Battye2002, Orlova2016}, merons \cite{Kasamatsu2004} and monopoles \cite{Stoof2001, Ruostekoski2003}.


In the present work we are interested in the few-vortex 2D dynamics for spinless, immiscible, mixtures of two components. Vortices in the majority component play the role of effective trapping potentials, which confine the minority component within the vortex cores.
Experimentally, a method for tracking the vortex dynamics was implemented in Ref. \cite{Kwon2021}, while, 
in a theoretical perspective, Lorentz-like equations were proposed for one or more vortices with massive cores \cite{Richaud2020,Richaud2022}. This picture was confirmed by applying the Lagrangian variational method to the Gross-Pitaevskii equations for a binary mixture with vortex excitations \cite{Richaud2021}, and was extended to the case of point-like masses independent from the positions of vortices
\cite{Bellettini2023}. 
In the latter works, the dynamics of vortices was characterized by considering the effects due to the presence of the second species.

In this work we consider the dynamics of a vortex pair (VP), where the vortices constitute an effective double well potential for the entrapped minority species $B$. We do not include the possible inter-well $B$-tunnelling. Unlike the previous works, we study the more general case of mass imbalanced vortices, where the total $B$ mass in the system can be differently distributed between the two vortices. This makes the dynamics of the two vortices much richer and closer to the experimental realizations of vortex pairs in binary condensates. 
To the best of our knowledge, a detailed analysis of the VP dynamics in such a case has not been carried out so far.

Here, based on the point-like vortex model discussed in Refs. \cite{Richaud2020,Richaud2022,Richaud2022,Richaud2023}, we  investigate mass imbalanced vortices by adopting the approximation that
the two $B$-masses are essentially concentrated in the vortex centers, with no significant fluctuations in their distribution between the vortices.
In such point-like framework, we find
a rich class of circular-orbit (CO) solutions for the imbalanced vortices (moving on constant-radius orbits), where the geometry of the two-vortex state significantly depends on the masses filling the vortex cores. We characterize them analytically and present the study of their linear stability.
Furthermore, we confirm the reliability of the analytical vortex trajectories by comparing them with numerical results, i.e. simulations of the VP dynamics as governed by the Gross-Pitaevskii equations for the mixture of Bose-Einstein condensates (BECs).

The space of parameters for two unbalanced, rotating vortices is five dimensional, and it gives rise to a plentiful scenario of CO solutions, which depend on the core masses, the two orbits radii and the angular velocity of the VP. The dynamical equations associated with the CO solutions offer then the possibility of playing with some model parameters, while inferring the others. In particular, our analysis highligths a viable scheme to infer the vortex masses (whose measurement is known to be a hard problem at an experimental level) from the VP geometry and the precession frequency.
In general, our work aims at providing a reliable model for experimental situations where the presence of vortices with imbalanced masses, rather than balanced masses, is expected.

The layout of the paper is as follows. First we introduce (section \ref{sec:Model}) the point-like (PL) model in the Lagrangian and Hamiltonian picture, along with our class of CO solutions. In section \ref{sec:Normal_modes} we apply the 
linear-stability analysis to our point-like vortex model. 
In section \ref{sec:frequency_branches} we investigate the rotational dynamics of the VPs and determine the significant properties of the precession frequency as a function of the masses and vortex radii. Additionally, we explore the stability character of the CO solutions in the small-oscillation regime. 
In section \ref{sec:mass_surfaces}
we adopt a different perspective that, by imposing the positive-mass condition, allows us to define the  diagram of the VP rotational states at a given precession frequency. This is leading to a classification of all the CO solutions, the determination of their properties and the highlight of the precession frequency role. 
In section \ref{sec:GP_results} we move to the Gross-Pitaevskii simulations and validate our analytic predictions. Lastly, in section \ref{sec:Conclusions} we present our final remarks and conclusions.

\section{Point-like vortex model}
\label{sec:Model}

The static and dynamical properties of a binary mixture of BECs are well described by two coupled Gross-Pitaevskii equations (GPE)
\begin{equation}
    i\hbar\frac{\partial}{\partial t}\psi_s=\bigg(-\frac{\hbar^2\nabla^2}{2 m_s}+V_{ext}(\bm{r})+g_s|\psi_s|^2+g_{sj}|\psi_j|^2\bigg)\psi_s,
    \label{gpe1}
\end{equation}
where $s,j\in \{A,B\}$ with $j\neq s$ and $m_A$, $m_B$ are the atomic masses of the two species. We assume that the vortices are nucleated in the majority component $A$, whereas $B$ does not host any defects.
The order parameters
$\psi_A$ and $\psi_B$ associated with the mixture components
are normalized such that the two conserved quantities:
$$
N_s = \int_D d^2 x \;|\psi_s|^2, \quad s = A, \;B, 
$$
represent the particles number of each species in the ambient space $D$. The intra-species and inter-species interactions
$g_s={4\pi\hbar^2 a_s}/{m_s}$ and
$g_{AB}=g_{BA}= {2\pi\hbar^2 a_{AB}}/{m_r}$
depend on the s-wave scattering lengths
$a_s$ and $a_{AB}$, respectively, and $m_r=(1/m_A+1/m_B)^{-1}$
is the reduced mass.
In the present paper, $g_{AB}$ and $g_s$ are positive, i.e. repulsive, and satisfy the condition $g_{AB}>\sqrt{g_A g_B}$ determining the immiscibility of the two species.
For simplicity, we treat here the case of a 2D trap where $D$ is a disk. Thus, $V_{ext}$ is a rigid-wall potential such that $V_{ext} = 0$ inside the disk. 
The effective 2D dynamics is experimentally reproduced via pancake-like potentials. To obtain this dimensional reduction, $g_s$ and $g_{AB}$ must be normalized, in the equations, by some effective condensate thickness $L_z$ (we always take: $L_z=2\times10^{-6}$ $m$).
In this $z$-direction, the dynamics is frozen. Accordingly, all the densities are planar densities.

The point-like model for vortices with 
massive cores in binary condensates
has been proposed in Ref. \cite{Richaud2020} and derived in Ref.
\cite{Richaud2021}, within a well-known Lagrangian variational
approach that was first used to investigate the dynamics of a Bose-Einstein condensate in a trap \cite{Zoller1996}. 
This variational procedure consists in first introducing a suitable ansatz for the order parameters $\psi_A$ and $\psi_B$. The ansatzes depend on the spatial coordinates and implicitly on time, via some chosen variables.
After substituting these ansatzes into the field Lagrangian relative to equations (\ref{gpe1}), one integrates out the field degrees of freedom in space, ending up with an effective Lagrangian that is a function of the vortex-core mass coordinates (chosen variational variables).
The resulting dynamical picture consists of Euler-Lagrange equations with a Lorentz-like form, which allow for a simple derivation of the CO solutions for a vortex pair. We also derive the corresponding Hamilton description, as it will prove particularly convenient for investigating the system’s stability properties.

\subsection{Point-like dynamics: from the Lagrangian to the Hamiltonian picture}
\label{sec:equations}
The effective Lagrangian describing the 2D dynamics of $N_v$ point-like vortices 
with massive cores reads:
\begin{equation}
    \mathcal{L}=
    \sum_{i=1}^{N_v} \left (\frac{1}{2}m_i\dot{\bm{r}}_i^2+ \kappa_i\frac{\rho_A}{2}\dot{\bm{r}}_i\wedge\bm{r}_i\cdot\hat{e}_3 
    \right ) -E(\bm{r}_1,...,\bm{r}_{N_v}),
\end{equation}
for a disk geometry with $R$ the radius of the disk. Each vortex center is assumed to be coincident with its core mass, in light of the immiscibility of the two components.
$m_i$, the core mass of the $i$th vortex, is a fraction of 
$M_B = m_B N_B$, the total $B$-component mass of the $N_B$ atoms.  
$N_A$ is the number of atoms of component $A$, $\rho_A =N_A m_A$ is its (planar) mass density and $n_A=\frac{N_A}{\pi R^2}$ its number density. The vortex-array energy, including the effects of boundary,
is
\begin{multline}
    E(\bm{r}_1,...,\bm{r}_{N_v}) = 
    \frac{ \rho_A}{4\pi} \sum_{i=1}^{N_v} 
    \kappa_i^2 \ln\left( 1-\frac{r_i^2}{R^2} \right) +\\
    +\frac{\rho_A}{4\pi}\sum_{i < j=1}^{N_v} \kappa_i \kappa_j \ln\left( \frac{
 R^4 -2R^2 \bm{r}_i \cdot\bm{r}_j +r_j^2 r_i^2 }{R^2|\bm{r}_i -\bm{r}_j |^2}\right),
 \label{eq:poten}
\end{multline}
where  $\kappa_i=N_i{h}/{m_A}$ is the $i$th vorticity charge 
and $N_i=\pm 1$. 
We now focus on the case of two vortices, $N_v=2$, contained in the 2D disc, where 
$\bm{r}_1=(x_1,y_1)$ and $\bm{r}_2=(x_2,y_2)$ are the vortex position vectors and $r_i<R$.
The Euler-Lagrange equations read
\begin{equation}
     m_\ell\Ddot{\bm{r}}_\ell = -\rho_A \kappa_\ell \dot{\bm{r}}_\ell 
     \wedge\hat{e}_3 - \nabla_{\bm{r}_\ell}E
     \label{eq:equations_of_motion}
\end{equation}
with
\begin{multline}
\!\! \nabla_{\bm{r}_\ell}E=    
\sum_{i\neq \ell }^{2}\frac{\rho_A \kappa_\ell \kappa_i}{2\pi}\left( \frac{ r_i^2\bm{r}_\ell 
-R^2\bm{r}_i}{ D(\bm{r}_i, \bm{r}_\ell )} 
- \frac{\bm{r}_\ell-\bm{r}_i}{ |\bm{r}_\ell - \bm{r}_i |^2 }\right)
    \label{eq:eom}
\end{multline}
and 
\begin{equation}
D(\bm{r}_i, \bm{r}_\ell ) \equiv R^4-2 R^2 \bm{r}_\ell \cdot\bm{r}_i+r_\ell^2r_i^2,
\label{eq:simbD}
\end{equation}
$l,i=1,2$, $l\ne i$.
These equations feature two conserved quantities. The first is the angular momentum of the planar system 
\begin{equation}
L_3= \sum_{i = 1 }^{2} \left (  -\frac{\kappa_i \rho_A}{2}r_i^2+m_i(\bm{r}_i\wedge\dot{\bm{r}}_i)\cdot\hat{e}_3
\right ).
\end{equation}
By defining the linear momenta
\begin{equation}
p_{xi} 
=m_i\dot{{x}}_i+\frac{\rho_A \kappa_i}{2} y_i ,
\quad
p_{yi} 
= m_i\dot{{y}}_i - \frac{\rho_A \kappa_i}{2} x_i,
\end{equation}
($i=1,2$) of the two vortices within the Lagrangian picture, $L_3$ takes the Hamiltonian form
\begin{equation}
L_3 = \sum_{i=1}^{2} (x_ip_{yi}-y_ip_{xi} ).
\end{equation}
The second conserved quantity is the total energy of the system, represented by the Hamiltonian. This is
\begin{equation}
    =  \sum_{i =1}^2 \left ( \frac{p_i^2}{2 m_i} 
    + \frac{\rho_A \kappa_i }{2 m_i} L_{3i}
    +
    \frac{\rho_A^2 \kappa_i^2}{8m_i} 
    r_i^2 \right ) 
    + U(\bm{r}_1, \bm{r}_2),
    \label{eq:Hamlab}
\end{equation}
where $L_{3i} =(\bm{r}_i\wedge\bm{p}_i)\cdot \hat{e}_3$,
and the potential $U(\bm{r}_1, \bm{r}_2)$ is obtained form formula \eqref{eq:poten} with $N_v =2$.
The components $x_i$, $y_i$ of the vector $\bm{r}_i$ and the relative momenta 
$p_{x i}$, $p_{y i}$, $i=1,2$, 
satisfy the following Poisson brackets:
 $$          \{A,B\} = \sum_{i=1}^2\bigg[\frac{\partial A}{\partial x_i} \frac{\partial B}{\partial p_{xi}} + \frac{\partial A}{\partial y_i} \frac{\partial B}{\partial p_{yi}} - \frac{\partial B}{\partial x_i} \frac{\partial A}{\partial p_{xi}} - \frac{\partial B}{\partial y_i} \frac{\partial A}{\partial p_{yi}}\bigg].
$$
The Hamilton equations relative to the Hamiltonian \eqref{eq:Hamlab} are derived in appendix \ref{sec:Hamilt}.

\subsection{Notable class of solutions: corotating imbalanced vortices}
\label{sec:circular_solutions}

The class of CO solutions for two massive vortices of same circulation is easily obtained 
with the simple ansatz 
\begin{equation}
    \bm{r}_1=\begin{pmatrix} x_1\\ y_1 \end{pmatrix}=\begin{pmatrix} a\;\cos{(\Omega t)}\\ a \; \sin{(\Omega t)} \end{pmatrix}\label{eq:ansatz_1}
\end{equation}
\begin{equation}
    \bm{r}_2=\begin{pmatrix} x_2\\ y_2 \end{pmatrix}=\begin{pmatrix} -b\;\cos{(\Omega t)}\\ -b \; \sin{(\Omega t)} \end{pmatrix}\label{eq:ansatz_2}
\end{equation}
where the two position vectors feature a constant angular shift of $\pi$, $a, b<R$, and $\Omega$ represents the angular velocity of the vortices along their circular orbits.
After plugging the ansatz above into the equations of motion \eqref{eq:equations_of_motion}, the final equations, relating the constant parameters $a$, $b$, $m_1$ and $m_2$, given $\Omega$, are:
\begin{multline}
    -m_1\Omega^2a+ \rho_A \kappa_1\Omega a-\frac{\rho_A \kappa_1^2}{2\pi}\frac{a}{R^2-a^2}+\\
    +\frac{\rho_A \kappa_1\kappa_2}{2\pi}\left[\frac{b^2a+bR^2}{(R^2+ab)^2}-\frac{1}{a+b}\right]=0 
\end{multline}
\begin{multline}
    m_2\Omega^2b-\rho_A \kappa_2\Omega b+\frac{\rho_A \kappa_2^2}{2\pi}\frac{b}{R^2-b^2}+\\
    +\frac{\rho_A \kappa_1\kappa_2}{2\pi}\left[\frac{-a^2b-aR^2}{(R^2+ab)^2}+ \frac{1}{b+a}\right]=0 
\end{multline}
If $\kappa_1=\kappa_2$ and $m_1=m_2$ the system becomes symmetric under the exchange of $a$ and $b$, allowing thus for the solution $a=b$. 
Hereafter, we assume that the vortices feature the same vorticity charge: $\kappa_1= \kappa_2 = \kappa>0$. 
In this case, the equations reduce to
\begin{equation}
(\rho_A \kappa - m_1\Omega)\Omega a - \frac{\rho_A \kappa^2}{2 \pi} F(a,b) =0,
\label{eq:paramEq1}
\end{equation}
\begin{equation} 
(\rho_A \kappa-m_2\Omega)\Omega b  - \frac{\rho_A \kappa^2}{2 \pi} F(b,a) =0,
\label{eq:paramEq2}
\end{equation}
where
\begin{equation} 
F(a,b) = \frac{1}{a+b} - \frac{b}{ab+R^2}+ \frac{a}{R^2-a^2}.
\label{eq:effe}
\end{equation}
Note that both $F(a,b)$ and $F(b,a)$ are always positive due to the condition $a,b < R$.
Such equations link $a$, $b$, $\Omega$, $m_1$ and $m_2$. The domain of the
physical values for $a$ and $b$ is determined by imposing the conditions 
\begin{equation}
m_1 \ge 0, \quad m_2 \ge 0,
\label{eq:m1m2}
\end{equation}
where $m_1 = m_1(a,b,\Omega)$ and  $m_2 = m_2(a,b,\Omega)$ are obtained from
equations \eqref{eq:paramEq1} and \eqref{eq:paramEq2}, respectively.

\section{small-oscillations}
\label{sec:Normal_modes}

We proceed with finding the excitation frequencies of the VP, as done in Ref. \cite{Zoller1996}.
These are given by the small-oscillation frequencies around the fixed points of the dynamical system corresponding to equations \eqref{eq:equations_of_motion}. 
To this purpose, we switch to the Hamiltonian formalism and go to a rotating frame of reference, where the CO solutions of subsection \ref{sec:circular_solutions} are fixed points of the new dynamical system.
After introducing the formalism, we characterize the stability/instability of the equilibrium points with respect to the low-energy excitations.

As mentioned, we consider a rotating frame of reference, with angular velocity $\Omega$. Here, our ansatz solutions are fixed points.
The Hamiltonian in the (primed) rotating system is
\begin{equation}
    \mathcal{H}_{rot}(\bm{z}') = \mathcal{H}(\bm{z}') - \Omega L_3(\bm{z}'),
    \label{eq:HamiltonianRot}
\end{equation}
where the vector of the dynamical variable $\bm{z}$ is
$\bm{z}'=(\bm{r}'_1,\bm{p}'_1,\bm{r}'_2,\bm{p}'_2)$ (rotating reference frame).
The Hamiltonian $\mathcal{H}_{rot}$ of the rotating system is in the primed variables, linked to those of the original system by the transformation matrix $R$,
\begin{equation}
     	R=\begin{pmatrix}
\cos{(\Omega t)} & -\sin{(\Omega t)} \\
\sin{(\Omega t)} & \cos{(\Omega t)}
\end{pmatrix}
\end{equation}
\begin{equation}
    \bm{r}_i= R\;\bm{r}'_i,
    \label{eq:coord_transf}
\end{equation}
where $\bm{r}_i$ and $\bm{r}'_i$ are the spatial coordinates of the vortices ($i=1,2$) respectively in the laboratory and in the rotating reference frame. 
By plugging the relation \eqref{eq:coord_transf} into equation \eqref{eq:Hamlab} (and changing the momenta accordingly), one obtains the rotating Hamiltonian \eqref{eq:HamiltonianRot}.
The Hamilton equations in the rotating system are:
\begin{equation}
    \dot{\bm{r}}_i'=\frac{\bm{p}_i'}{m_i}+\frac{\kappa\rho_A}{2 m_i} \hat{e}_3\wedge \bm{r}_i'+\Omega \bm{r}'_i\wedge\hat{e}_3,
    \label{eq:Ham_eqs_ri_prime}
\end{equation}

\begin{equation}
        \dot{\bm{p}}_i' = 
        \left( \frac{\kappa\rho_A}{2 m_i} -\Omega
        \right ) \hat{e}_3\wedge\bm{p}_i'
        -\nabla_{\bm{r}_i'} U(\bm{r}_1',\bm{r}_2')
        -\frac{\kappa^2 \rho_A^2 \bm{r}_i'}{4 m_i}
        ,
        \label{eq:Ham_eqs_pi_prime}
\end{equation}
where
\begin{multline}
    -\nabla_{\bm{r}_i'} U(\bm{r}_i',\bm{r}_j') = 
     \frac{\kappa^2\rho_A \bm{r}_i'}{2 \pi (R^2 -r_i'^2)} + 
    \\+\frac{\kappa^2 \rho_A}{4 \pi }\left(
    \frac{ 2R^2\bm{r}_j' + r_j'^2\bm{r}_i' }{
    D(\bm{r}_i', \bm{r}_\ell')}  
    +2\frac{\bm{r}_i' - \bm{r}_j'}{(\bm{r}_i'-\bm{r}_j')^2} \right),
\end{multline}
and the symbol $D(\bm{r}_i', \bm{r}_\ell' )$ is defined by equation (\ref{eq:simbD}). By introducing the vector 
$$\bm{z} = (x_1, y_1, p_{1,x}, p_{1,y}, x_2, y_2, p_{2,x}, p_{2,y})^T
$$ 
(for the sake of simplicity, we remove the symbol $'$ from the Hamiltonian variables)
equations (\ref{eq:Ham_eqs_ri_prime}) and (\ref{eq:Ham_eqs_pi_prime}) can be expressed in the more compact form 
$ \dot{\bm{z}} = \mathbb{E} \nabla_{\bm{z}} \mathcal{H}_{rot} $,
where $\mathbb{E}$ is the
symplectic matrix
\begin{equation}
    \mathbb{E}=\left(
\begin{array}{cccccccc}
 0& 0& 1& 0& 0& 0& 0& 0 \\
 0& 0& 0& 1& 0& 0& 0& 0 \\
 -1& 0& 0& 0& 0& 0& 0& 0 \\
0& -1& 0& 0& 0& 0& 0& 0 \\
 0& 0& 0& 0& 0& 0& 1& 0 \\
 0& 0& 0& 0& 0& 0& 0& 1 \\
 0& 0& 0& 0& -1& 0& 0& 0 \\
 0& 0& 0& 0& 0& -1& 0& 0
\end{array}
\right)
\end{equation}
After expanding around the equilibrium points of the dynamical system, and linearizing, we obtain:
\begin{equation}
    \dot{\bm{\xi}} = \mathbb{J} \bm{\xi}, 
    \label{eq:dyn_system}
\end{equation}
where the Jacobian matrix is $\mathbb{J}=\mathbb{E}\mathbb{H}$, and $\bm{\xi}$ is the vector of the perturbations of the variables $\bm{z}$ (see Appendix \ref{appendix:small_oscillations}).
$\mathbb{H}$ is the symmetric Hessian matrix of $\mathcal{H}_{rot}$, computed in $\bar{\bm{z}}$, the fixed point of the dynamical system (see Supplemental Material). The eigenvalues of $\mathbb{J}$ determine the stability character of the equilibrium. In particular, whenever an eigenvalue features a non-zero real part, the equilibrium is unstable (for later purposes: $Re(\lambda_i)$, $\lambda_i$ eigenvalue of $\mathbb{J}$, is considered zero if of the order of $10^{-7}$ or smaller).

\section{Rotational dynamics of the vortex pairs}
\label{sec:frequency_branches}
A first characterization of the rotational dynamics of the VP, described by the CO solutions, is found by analyzing the precession frequency of the vortices $\Omega$. In this picture, $\Omega$ is studied at varying mass of one of the vortices,
for a given choice of the radii $a$ and $b$. 
We solve equation \eqref{eq:paramEq1} to obtain $\Omega$ as a function of $m_1$ while Eq. \eqref{eq:paramEq2}, once $\Omega$ is known, determines $m_2$. The frequency features two branches
\begin{equation}
    \Omega_1 (m_1,a,b) =\frac{\kappa \rho_A}{2m_1 } \bigg( 1 
    -\sqrt{ 1-\frac{2 m_1}{\pi a \rho_A} F(a,b) }\bigg),
\label{eq:ome1}
\end{equation}
and
\begin{equation}
    \Omega_2 (m_1,a,b) =\frac{\kappa \rho_A}{2m_1 } \bigg( 1 
    +\sqrt{ 1- \frac{2 m_1}{\pi a \rho_A} F(a,b) }\bigg),
\label{eq:ome2}
\end{equation}
\begin{figure}
    \centering
    \includegraphics[width=0.43\textwidth]{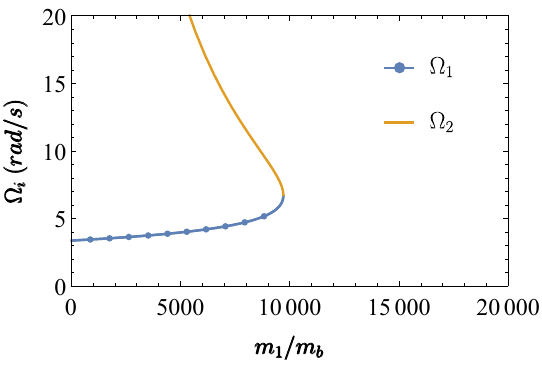}
    \caption{Once vortex radial positions $a$ and $b$ are set, Eqs. \eqref{eq:ome1} and
    \eqref{eq:ome2} provide the two branches of the precession frequency $\Omega$, at varying $m_1/m_B$. $N_A=10^5$, $a/R=0.7$, $b/R=0.3$.}
    \label{fig:omega_Branches_vs_m1}
\end{figure}
\begin{figure}
         \centering
         \includegraphics[width=0.45\textwidth]{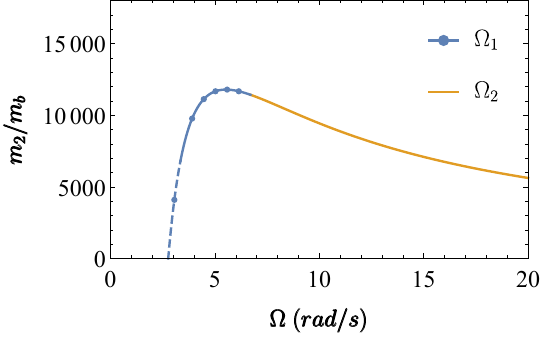}
         \caption{After finding $\Omega$, as in Fig. \ref{fig:omega_Branches_vs_m1}, we subsequently find $m_2/m_B$ as a function of $\Omega$. The dashed line represents the region $\Omega < \Omega_{1m}$ where $\Omega$ satisfies equation \eqref{eq:paramEq2} but not \eqref{eq:paramEq1}. $N_A=10^5$, $a/R =0.7$, $b/R=0.3$.}
        \label{fig:m2_vs_omega_Branches}
\end{figure}
\medskip

\noindent
with $F$ defined by (\ref{eq:effe}), having physical meaning if the inequality
$\pi a\rho_A \ge 2 m_1 F(a,b)$ is satisfied. 
The same inequality determines the range of $m_1$ 
\begin{equation}
\frac{\pi a \rho_A}{2F(a,b)} \ge  m_1
\label{dom1}
\end{equation}
as a function of the density $\rho_A$ and of the CO-solution radii (note that $F(a,b)$ can be shown to be always strictly positive.). The maximum value of $m_1$, well visible in Fig. \ref{fig:omega_Branches_vs_m1}, is the point at which 
$\Omega_1 \equiv \Omega_2$. In the limit $m_1 \to 0$, the frequency $\Omega_1 (m_1,a,b)$ tends to the minimum value
\begin{equation}
\Omega_{1m} =
\Omega_1 (0,a,b) =
\frac{ \kappa}{2 \pi a} F(a,b).
\label{ome1m}
\end{equation}
This describes the rotation of a vortex pair where vortex 1 is massless while $m_2 \ne 0$. In this limit,
$\Omega_2 (m_1,a,b)$ provides instead an unphysical case since $\Omega_2$ diverges for $m_1 \to 0$.

An equivalent approach consists in solving Eq.
\eqref{eq:paramEq2} at given $a$, $b$ and $m_2$, to then determine $m_1$ via Eq. \eqref{eq:paramEq1}.
This scheme, based on Eq. \eqref{eq:paramEq2}, provides the frequencies
\begin{equation}
\Omega_1' =\Omega_1 (m_2,b,a), \quad 
\Omega_2' =\Omega_2 (m_2,b,a),
\label{ome12}
\end{equation}
obtained by replacing $m_1$, $a$ and $b$ with $m_2$, $b$ and $a$, respectively, in equations (\ref{eq:ome1}), and (\ref{eq:ome2}). The different parametrization, involving $m_2$, highlights some new information about the physical range of $\Omega$: while $m_2 \to 0$ entails the unphysical situation $\Omega_2' \to \infty$
(see Fig. \ref{fig:m2_vs_omega_Branches}), for $m_2 \to 0$ the frequency $\Omega_1'$ features a minimum value, that is
\begin{equation}
\Omega_{1m}'
=\Omega_1 (0,b,a) =
\frac{ \kappa}{2 \pi b} F(b,a) 
\ne \Omega_{1m} .
\label{ome1mp}    
\end{equation}
The lowest possible value for $\Omega$ is thus determined by
the largest value between the quantities $\Omega_{1m}'$ and $\Omega_{1m}$, emerging from the two approaches.

Summarizing,  the range of $\Omega$, as well as the ranges of $m_1$ and $m_2$, are related by the fact that the mass $m_1$ ($m_2$) can tend to zero, implying the frequency range $\Omega \in [\Omega_{1m} , \infty]$
($\Omega \in [\Omega_{1m}' , \infty]$),
while the other mass $m_2$ ($m_1$) tends to a finite value in correspondence to the minimum value $\Omega= \Omega_{1m}$ ($\Omega = \Omega_{1m}'$)
if 
$$
\Omega_{1m}' < \Omega_{1m} \quad
\bigl (\Omega_{1m} < \Omega_{1m}' \bigr ).
$$
The choice between the two approaches, which can be shown to be perfectly equivalent,  depends on the constants $a$ and $b$. In the case of $a=b$ one easily proves that
$\Omega_{1m}' \equiv \Omega_{1m}$, as $F(a,b)= F(b,a)$.

\begin{figure}
    \centering         
    \includegraphics[width=0.45\textwidth]{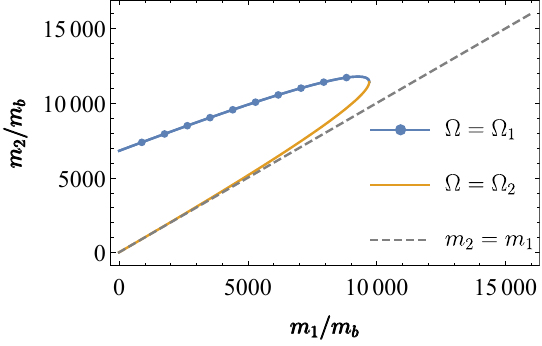}
    \caption{Plot of $m_2/m_B$ vs $m_1/m_B$ at $a/R=0.7$, $b/R=0.3$, $N_A=10^5$. This is obtained by combining
    $m_2(\Omega,a,b)$
    (obtained from Eq. \eqref{eq:paramEq2})
    either with 
    $\Omega= \Omega_1(m_1,a,b)$
    (blue dotted branch),
    or with
    $\Omega= \Omega_2 (m_1,a,b)$
    (orange dashed branch). $\Omega_1$ and
    $\Omega_2$ are given by 
    Eqs. \eqref{eq:ome1} and \eqref{eq:ome2},
    respectively. When $m_1 \to 0$ then $m_2$ tends to a finite value.}
    \label{fig:m2_vs_m1_2}
\end{figure}

\begin{figure}
     \centering
     \includegraphics[width=0.45\textwidth]{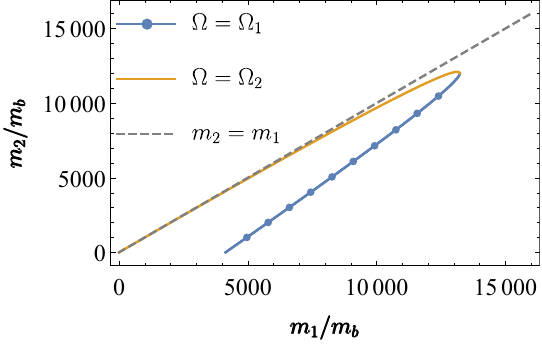}
         \caption{ This plot at $a/R=0.5$, $b/R=0.6$, and 
         $N_A=10^5$ describes $m_2/m_B$ as a function of $m_1/m_B$ within the second scheme ($\Omega_{1m}<\Omega_{1m}'$). In this case $m_1 \ne 0$ for $m_2 \to0$.}
         \label{fig:m2_vs_m1_1}
\end{figure}

Let us analize the interplay  among $\Omega$, $m_1$, and $m_2$ within the first approach, by considering the specific case $a=0.7R$ and $b=0.3R$. For this choice the limit $m_1 \to 0$ is allowed. 
Fig. \ref{fig:omega_Branches_vs_m1} describes {\it i}) the two branches of frequency $\Omega_1(m_1,a,b)$ and $\Omega_2(m_1,a,b)$ in different colors, {\it ii}) the frequency $\Omega_{1m}$ for $m_1\to 0$ (see Eq. \eqref{ome1mp}), and {\it iii}) the maximum value of $m_1$,
\begin{equation}
m_{1,max} = \frac{ \pi a \rho_A}{2F(a,b)} ,
\label{eq:m1max}
\end{equation}
predicted by Eq. (\ref{dom1}) at $\Omega = \kappa F(a,b)/(\pi a)$. In this figure (and in the following ones) we use the dimensionless quantities $m_i/m_B$, $i=1,2$, where $m_B$ is the mass of a species-$B$ atom.
Fig. \ref{fig:m2_vs_omega_Branches} displays, for $a$ and $b$ as in Fig. \ref{fig:omega_Branches_vs_m1}, the profile of the mass $m_2(\Omega, a,b)$ of the second vortex,
given by Eq. \eqref{eq:paramEq2}. The maximum value
\begin{equation}
m_{2,max}= \frac{ \pi b \rho_A}{2F(b,a)}
\label{eq:m2max}
\end{equation}
is reached at $\Omega = \kappa F(b,a)/(\pi b)$.
Note that the chosen radii $a$ and $b$ are such that $\Omega_{1m}' < \Omega_{1m}$, implying that the range of $\Omega$ is $[\Omega_{1m}, \infty]$. In the limit $\Omega\to\Omega_{1m}$, one finds the expected limit $m_1\to 0$, whereas $m_2 (\Omega_{1m}, a, b)$ is finite and its value separates the dashed from the dotted (blue) arc in Fig. \ref{fig:m2_vs_omega_Branches}. This situation is also represented in Fig. \ref{fig:m2_vs_m1_2}, where $m_2$ is expressed as a function of $m_1$ for $a=0.7R$, $b=0.3R$. 
Fig. \ref{fig:m2_vs_m1_1} illustrates instead the opposite behavior, at $a=0.6R$, $b=0.6R$, where $m_1$ achieves a finite value while $m_2 \to 0$. Such limit cases highlight the class of CO solutions, with $\Omega$ given by \eqref{ome1m} and \eqref{ome1mp}, where only one vortex is massless and $a\neq b$.

The limit where both masses go to zero, represented by the branch $\Omega_2$, is always present (see Figs. \ref{fig:omega_Branches_vs_m1} and \ref{fig:m2_vs_omega_Branches}). The corresponding CO solution is however generally unphysical, for it involves high or diverging frequencies.
Very interestingly, solutions where both vortices are massless are predicted by equations (\ref{eq:ome1}), and (\ref{eq:ome2})
when the condition $\Omega_1(0, a,b) = \Omega_1'(0, b,a)$ is imposed (see \eqref{ome1m} and \eqref{ome1mp}).
Such more complicated situations will be discussed in section \ref{sec:mass_surfaces}.

\begin{figure}
    \centering
    \includegraphics[width=0.35\textwidth]{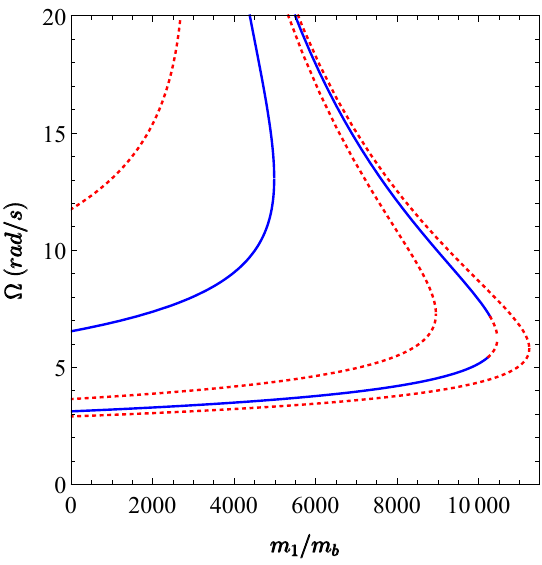}
    \caption{Precession frequency $\Omega$ as a function of the first vortex mass $m_1$, described by
    Eqs. (\ref{eq:ome1}) and (\ref{eq:ome2}). 
We show curves at  (from the left to the right) $a/R=0.95$, $0.20$, $0.80$, $0.42$ and $0.7$, with $b/R=0.5$, $N_A=10^5$. Blue continuous (red, dotted) arcs highlight the stable (unstable) behaviors.}
    \label{fig:omega_branches_stab}
\end{figure}

\subsection{Stability character of CO solutions}
\label{sec:stability}

\begin{figure}
\includegraphics[width=0.35\textwidth]{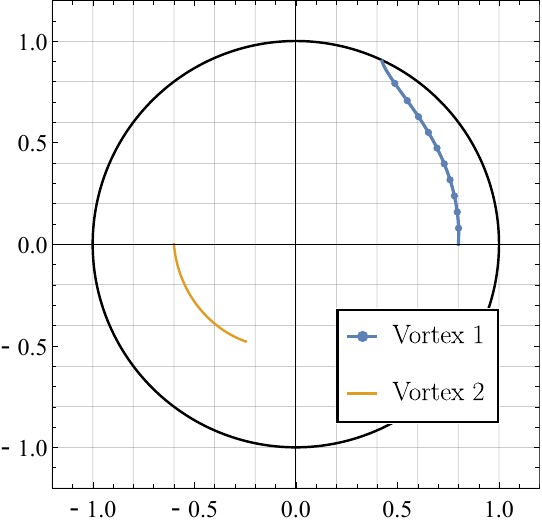}
\caption{Unstable CO solution with $a/R=0.8$, $b/R=0.6$, $N_1\simeq 8300$, $N_2\simeq 15100$, $\Omega=5.26$ $rad/s$, and $N_A=10^5$. The vortex trajectories are caused by a small perturbation of initial position $(a,0)$. Running time: $t_{max}\simeq 0.2$ $s$, such that vortex $1$ hits the boundary.}
\label{fig:evident_instability}
\end{figure}

The application of the stability analysis discussed in Sec. \ref{sec:Normal_modes} to the class of CO solutions allows us to determine the stable or unstable character of the VP motion. This property is illustrated in the plot of $\Omega$ as a function of $m_1$ (as the one represented in Fig. \ref{fig:omega_Branches_vs_m1}) for a given $b$ and different values of $a$. Fig. \ref{fig:omega_branches_stab} shows the curves corresponding to $a/R = 0.20$, $0.42$, $0.70$, $0.80$ and $0.95$, for $b/R= 0.5$. Blue continuous curves and red dashed curves are associated with a stable and unstable behavior, respectively. Hybrid curves can be found where stability arcs intercalate instability arcs. This is the case of the curve at $a/R=0.42$, where most of the values of $m_1$ correspond to a regime characterized by bistability (two stable solutions for each value of $m_1$), whereas, for $m_1$ large enough, one can intercept a frequency interval exhibiting instability.
Note that the increasing sequence $a/R = 0.20$, $0.42$, $0.70$, $0.80$ and $0.95$ does not correspond to the sequence of curves, from left to right, shown in Fig. \ref{fig:omega_branches_stab}. This effect is explained in section \ref{sec:mass_surfaces}
where the complex interplay between the masses ($m_1$ in this case) and the orbit radii $a$ and $b$ is investigated.
Fig. \ref{fig:evident_instability} represents an example of instability manifestation. Upon a
small perturbation, the trajectory of vortex 1 gets increasingly
farther from its initial position $(a, 0)$
and collides with the boundary.

{\it Eigenmodes of a balanced system}. To get some insights into the stability regimes of the VP, we consider the small (stable) 
oscillations around a CO solution in the simplified scenario where $a=b$ and $m_1=m_2$. In this case, the system features three pairs $\pm\omega_i$, $i=1,2,3$ of eigenfrequencies that are not zero. They reflect three different oscillations modes, which can be triggered close to a fixed point. As mentioned, hereafter we restrict ourselves to the case of a symmetric VP, a state for which $\omega_1=\omega_2$. In Appendix \ref{sec:appendix_eigenmodes} the eigenmodes are discussed for the more general case of a mass-imbalanced VP.

\begin{figure}
    \centering
    \includegraphics[width=0.35\textwidth]{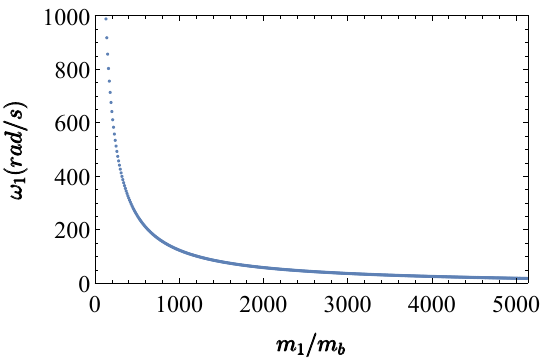}
    \caption{First eigenfrequency of the VP as a function of the vortex-1 mass for $a/R=b/R=0.3$, $m_1=m_2$, and $N_A=10^5$. This mainly represents the radial oscillations of the two vortices.
    The higher the masses, the slower are the two vortices, in accordance with \cite{Bellettini2023}. For any value of $m_1$, we take the corresponding $\Omega_1$ (see Eq. \eqref{eq:ome1}) for the global precession frequency and thus compute the eigenfrequency $\omega_1$ of the dynamical system.}
    \label{fig:radial_osc_freq_balanced_vs_m1}
\end{figure}

\begin{figure}
    \centering   
    \includegraphics[width=0.4\textwidth]{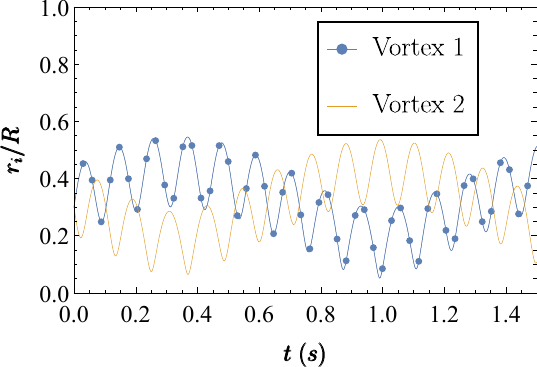}
    \caption{Under identical perturbations on the vortex initial velocities, some $\pi$-phase shifted radial oscillations (described by radii $r_i$, $i=1,2$) are triggered which can be associated with eigefrequency $\omega_2$. These are superimposed to larger time-scale oscillations associated with $\omega_3$. $N_A=10^5$, $a/R=b/R=0.3$, $N_1=N_2=2000$.}
    \label{fig:balanced_case_radial_phase_shifted}
\end{figure}

Fig. \ref{fig:radial_osc_freq_balanced_vs_m1} shows the first eigenfrequency $\omega_1$ of the dynamical equations \eqref{eq:dyn_system}, as a function of the vortex masses, for balanced vortices within the stability regime. 
The eigenmode of frequency $\omega_1$ is mainly associated with in-phase radial oscillations of the VP (see \cite{Bellettini2023}). Note that every point of the plot in Fig. \ref{fig:radial_osc_freq_balanced_vs_m1}
also implies a different precession frequency $\Omega$ and a different $m_1$ ($ =m_2$), satisfying equations \eqref{eq:paramEq1} and \eqref{eq:paramEq2}. 
The higher the masses, the slower are the two vortices, in accordance with \cite{Bellettini2023}.
The in-phase radial oscillations of the two vortices can be shown to be accompanied by no motion of the centre of mass. 
\\An opposite behavior, on the other hand, characterizes the eigenmode relevant to $\omega_2$ which exhibits $\pi$-phase shifted radial oscillations of the two vortices, combined with a net motion of the centre of mass. These are shown in Fig. \ref{fig:balanced_case_radial_phase_shifted}, where the oscillations associated with $\omega_2$ are superimposed to the slower oscillations associated with the third frequency $\omega_3$.

Lastly, the eigenmode associated with $\omega_3$, analogously to the second eigenmode $\omega_2$, mainly describes $\pi$-phase shifted radial oscillations of the two vortices that have a much larger period (see Fig. \ref{fig:balanced_case_radial_phase_shifted}). This slow variation of the orbit radius is again associated with a net motion of the centre of mass of the symmetric VP. 

Remarkably, while the first two types of small-oscillations (associated to $\omega_1$ and $\omega_2$) vanish at small vortex masses (infinite frequency), the eigenmode associated with $\omega_3$ acquires a finite oscillation frequency. Conversely, at medium/high vortex masses, evident radial oscillations arise as a signature of the filled core. 

\section{Diagram of rotational states of vortex pairs}
\label{sec:mass_surfaces}

\begin{figure}
         \centering
         \includegraphics[width=0.25\textwidth]{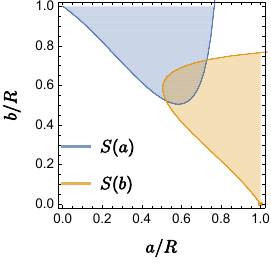}
         \caption{Shaded areas in the planar box $[0,1]\times[0,1]$ where $m_1(a,b)>0$ (blue) and $m_2(a,b)>$ $0$ (orange). 
         Curve (\ref{eq:ab}) ((\ref{eq:ba})) (zero-mass loci) represents the boundaries of the orange (blue) region. 
        The intersection of such regions defines the domain $\cal D$ of all the possible CO solutions, at $\Omega = 2.68$ $rad/s$ and
        $N_A=10^5$.
        }
         \label{fig:two_intersections}
\end{figure}

\begin{figure}
         \centering
         \includegraphics[width=0.25\textwidth]{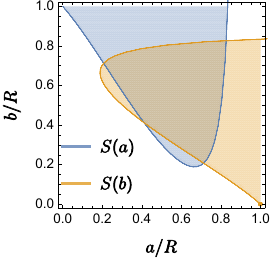}
         \caption{The same plot as in Fig. \ref{fig:two_intersections}, with $\Omega = 3.65$ $rad/s$.
         }
         \label{fig:four_intersections}
\end{figure}

\begin{figure}
         \centering
         \includegraphics[width=0.25\textwidth]{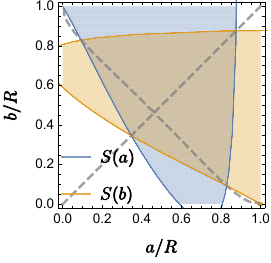}
         \caption{The plot is analogous to that shown in Fig. \ref{fig:two_intersections}, with: $\Omega =4.86$ $rad/s$, $q=4.40$. Dashed curves represent the pairs $(a,b)$ for which VPs exhibit equal masses.}
         \label{fig:four_intersections_cut}
\end{figure}

The CO solutions are well defined when inequalities \eqref{eq:m1m2}
are satisfied. Given $\Omega$, the curves 
corresponding to the limit cases $m_1=0$ and $m_2=0$,
in the quadrant $\{(a,b): a, b \ge 0 \}$, allow for the determination of the domain of validity of this class of solutions. By rewriting equations \eqref{eq:paramEq1} and \eqref{eq:paramEq2} in the form
\begin{equation} 
  m_1(a,b)= \frac{\rho_A \kappa}{\Omega} 
  - \frac{\rho_A \kappa^2}{2\pi \Omega^2 a} F(a,b) 
    \label{eq:mass1}
\end{equation}
\begin{equation}
  m_2 (a,b)= \frac{\rho_A \kappa}{\Omega} 
  - \frac{\rho_A \kappa^2}{2\pi \Omega^2 b} F(b,a) 
    \label{eq:mass2}
\end{equation}
and setting $m_1=m_2=0$, 
one obtains the two symmetric (with respect to the exchange $a \leftrightarrow b$) curves
\begin{equation}
b= S(a) = \frac{
P(a)- (R^2+ a^2) G(a))}{2(1 +a G(a))},
\label{eq:ba}
\end{equation}
\begin{equation}
a= S(b)=
\frac{
P(b)-(R^2+b^2) G(b))}{2(1 +b G(b))},
\label{eq:ab}
\end{equation}
from equations (\ref{eq:mass1}) and (\ref{eq:mass2}), respectively, where
$P(x) = \sqrt{ 4R^2 +
G^2(x) (R^2 -x^2)^2}$, with
$x= a,b$, and
$$
G(x) =
x \left 
(\frac{q}{R^2}- \frac{1}{R^2-x^2}
\right ), \quad
q=\frac{2\pi \Omega R^2}{\kappa}.
$$
The term $G(x)$ is responsible for the dependence of $b=S(a)$ and $a=S(b)$ on the frequency (the definition interval of $\Omega$ is discussed in Supplemental Materials). Curves (\ref{eq:ba}) and (\ref{eq:ab}) describe the boundaries of the orange domains ($a \ge S(b)$) and blue domains ($b \ge S(a)$), respectively, illustrated in Figs. \ref{fig:two_intersections}, \ref{fig:four_intersections}, and \ref{fig:four_intersections_cut}.
Here, the shaded areas shows where $m_1(a,b)>0$ (blue) or $m_2(a,b)>0$ (orange). Their intersection
defines the domain $\cal D$ (i.e. the small leaf well visible in Fig. \ref{fig:two_intersections}) of physical solutions.

Each pairs $(a,b)$ inside $\cal D$ describes VPs with vortex cores equipped with non-zero masses $m_1$, $m_2$. Note that each solution (associated with) $(a,b)$ and masses $m_1$, $m_2$ is paired to the solution $(b,a)$ with exchanged masses due to the symmetric form of Eqs. (\ref{eq:mass1}) and (\ref{eq:mass2}) with respect to the diagonal $a=b$.
In the white, blue or orange regions outside $\cal D$ at least one of conditions $m_1 \ge 0$ and $m_2 \ge 0$ is violated leading to unphysical regimes. 

The domain $\cal D$ represents the $\Omega$-dependent diagram of the rotational states and is characterized by different types  of CO solutions.
Fig. \ref{fig:two_intersections}, for example, exhibits a subclass of solutions representing {\it single-mass VPs} for pairs $(a,b)$ associated with the points of the blue, lower arc ($a>b$) or the orange, upper arc ($b>a$) forming the boundary $\Gamma$ of $\cal D$. These two arcs are identified by the conditions
$$
b = S(a) \,\, (m_1 =0), 
\qquad 
a > S(b) \,\, (m_2 >0)
$$
and
$$
a = S(b) \,\, (m_2 =0), 
\qquad
b > S(a) \,\, (m_1 >0),
$$ 
respectively.

Interestingly, the two arcs forming $\Gamma$,
placed symmetrically with respect to the diagonal $a=b$ in Fig.
\ref{fig:two_intersections}, become four in Figs. \ref{fig:four_intersections}, and \ref{fig:four_intersections_cut}. This macroscopic change of the boundary structure appears for $\Omega$ large enough. At increasing $\Omega$, in fact, the lobe $\cal D$ broadens to eventually cover the whole box $\{(a,b): 0 \le a, b \le R \}$, while, in parallel, $\Gamma$ changes its structure. In particular, numerical simulations allows us to identify the critical value $\Omega_c$ for which, when
$$
\Omega > \Omega_c \simeq 3.1 \;\; rad/s,
$$
the intersection points of curves $a=S(b)$ and $b= S(a)$ (where both $m_1$ and $m_2$ are zero) change from two to four (see Figs. \ref{fig:four_intersections} and \ref{fig:four_intersections_cut}): In additon to the extremes of the diagonal ($a=b$) segment in $\cal D$, the zero-mass condition is then satisfied by two additional symmetric points, for which $a\neq b$.
Such four points characterize the new structure of the boundary $\Gamma$, where they correspond to the extremes of the four arcs forming $\Gamma$, well visible in Figs. \ref{fig:four_intersections} and \ref{fig:four_intersections_cut}.
As discussed above, excluding the extremes, the pairs $(a,b)$ along the arcs describe {\it single-mass VPs}. 

CO solutions representing equal-mass VPs ($m_1=m_2$) provide another interesting case. This class is characterized by the equation 
\begin{equation}
\frac{ 1}{a} F(a,b)
=
\frac{1}{b} F(b,a),
\label{fab}
\end{equation}
obtained from
Eqs. (\ref{eq:mass1}) and (\ref{eq:mass2}), when $m_1=m_2$ is assumed. For $\Omega > \Omega_c$, the solution of equation \eqref{fab} features two branches, illustrated in Fig. \ref{fig:four_intersections_cut} by the two dashed lines intersecting the four vertices of the domain $\cal D$. At the vertices, it holds $m_1=m_2=0$. The presence of equal-mass VPs with $a\ne b$ (transverse branch) supplies a further signature of the $\Omega > \Omega_c$ regime.
The transition to the regime $\Omega < \Omega_c$, in fact, removes the transverse branch, restricting equal-mass CO solutions to the diagonal segment in $\cal D$ where $a=b$. Fig.
\ref{fig:two_intersections}
describes this case.

For $\Omega < \Omega_c$, decreasing sufficiently $\Omega$ reveals the presence of a minimum value below which the blue and the orange regions do not overlap and the CO solutions no longer exist, as the inequalities \eqref{eq:m1m2} cannot be satisfied simultaneously. For $a=b$ and $m_1=m_2=0$, formulas \eqref{eq:mass1} and \eqref{eq:mass2} give 
$$
\Omega = \frac{\kappa}{2\pi a} F(a,a)
= \frac{\kappa}{2\pi }
\left(
\frac{1}{2a^2} +\frac{2a^2}{R^4 -a^4}
\right),
$$
whose solution determines the coordinates of the two intersection points of $a=S(b)$ and $b=S(a)$ along the diagonal $b=a$ of the quadrant. On the other hand, the fact that $\Omega \to \infty$ for $a \to 0$ and $a \to R$ highlights the presence of a minimum value $\Omega_0$ in the interval $a\in [0,R]$. The condition $d\Omega/da = 0$ gives
$$
\Omega_0 = 
\frac{\kappa}{4\pi R^2}
{3^{3/4}}{\sqrt {2+ \sqrt 3 }},
$$ 
which is the value for which the overlap of the blue and orange regions in figures
(\ref{fig:two_intersections})-(\ref{fig:four_intersections_cut}) reduces to a single point. Frequency $\Omega_0$ represents the lowest possible value of the VP rotational frequency in the class of CO solutions.

By remaining on the segment $a=b$ of $\cal D$, one can get further information about CO solutions with $m_1 =m_2\ne 0$, at a given $\Omega$. Eqs. (\ref{eq:mass1}) and (\ref{eq:mass2})
supply the equation
\begin{equation}
m =
\frac{\rho_A \kappa}{\Omega} \left ( 1 - \frac{\kappa}{2\pi \Omega a}
F(a,a) \right )
\label{eq:meq}
\end{equation}
($m=m_1$) describing {\it equal-mass} CO solutions for any frequency $\Omega$. The calculation of $dm/da =0$ identifies the maximum mass 
\begin{equation}
\bar m 
= \frac{\rho_A \kappa}{\Omega} \left ( 1 - \frac{3^{3/4} \kappa}{4\pi \Omega R^2}{\sqrt {2+ \sqrt 3 }}\right ),
\label{eq:mmax}
\end{equation}
for $\bar a = R [(2 - \sqrt 3)/3]^{1/4}$,
along the diagonal segment of $\cal D$ where $m \ge 0$. Its approximate extension
${R}/{\sqrt{2q}}
\le a \le 
R\sqrt{{(q-1)}/{q} }$, with $q=2\pi\Omega R^2/\kappa$,
is easily evinced from Eq. (\ref{eq:meq}). Note that $\bar m$ is the maximum mass value in the class of equal-mass CO solutions with $a=b$ when $\Omega <\Omega_c$, while for $\Omega >\Omega_c$ such value 
corresponds to the crossing point of the two equal-mass branches.

\begin{figure}
\centering
    \includegraphics[width=0.45\textwidth]{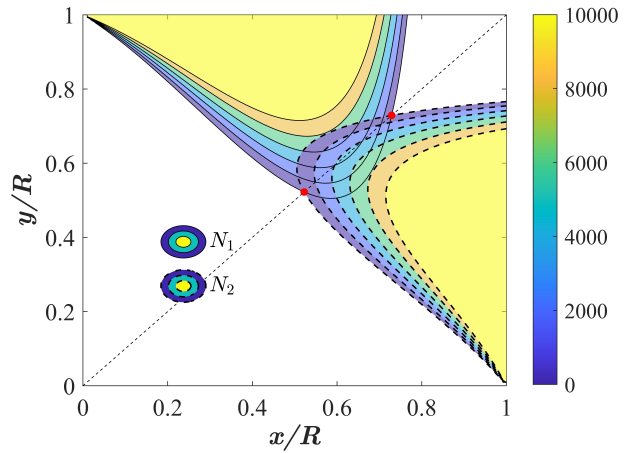}
\caption{\label{fig:masses_vs_a_vs_b} Number of $b$-particles $N_i=m_i/m_B$, with $i=1,2$, filling the VP cores as a function of $a$ and $b$, at $\Omega \simeq 2.68$ $rad/s$ and $N_A=10^5$. Here, only the mass range $N_i \in [0,10^4]$ is shown, namely only the values corresponding to  physically meaningful points are coloured. 
The red dots represent two massless vortices solutions at the chosen $\Omega$.}
\end{figure}

The contourplot, at fixed $\Omega$, of the function $m_1(a,b)/m_B$ and $m_2(a,b)/m_B$ 
that is shown in Fig. \ref{fig:masses_vs_a_vs_b} allows one to better visualize the previous scenario. As visible in figure, excluding the unphysical regions entailing at least one negative mass, a small ``leaf" remains, where the two surfaces are superimposed from above. This is the region $\cal D$ in Fig. \ref{fig:two_intersections}, where for any $a$, and $b$ both masses $m_1$ and $m_2$ are positive. In Fig. \ref{fig:masses_vs_a_vs_b}, the boundaries of the leaf 
are given by Eqs. \eqref{eq:ba} and \eqref{eq:ab}, 
while the two extreme points (the red points lying on the diagonal $a=b$ in the figure) are the two massless solutions at the chosen frequency $\Omega$. 
Unlike the regime with $\Omega > \Omega_c$, where two branches correspond to the locus of intersection of the two surfaces $m_1 (a,b)=m_2 (a,b)$, in the current case $\Omega = 2.63 \; rad/s < \Omega_c$, implying that the surface intersection only takes place on the diagonal $a=b$, with the mass profile defined by Eq. \eqref{eq:meq}. This
 resembles a concave parabola placed along the diagonal segment ($a=b$) of $\cal D$, suggesting that each value of $m$ can be associated with a pair of CO solutions with different radius $a$.

\begin{figure}
         \centering
         \includegraphics[width=0.25\textwidth]{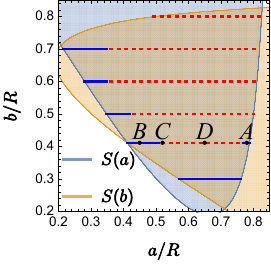}
         \caption{Same plot as in Fig. \ref{fig:four_intersections} (zoomed in), where the blue continuous segments (red dotted segments) represent $(a, b)$ pairs that feature dynamical stability (instability) for the relevant CO solutions (see section \ref{sec:stability}). Horizontal lines are drawn at: $b/R = 0.80$, $0.70$, $0.60$, $0.50$, $0.41$, $0.30$. The points A, B, C and D correspond to those in Fig. \ref{fig:omega_Branches_vs_m1_stab}.}
         \label{fig:four_intersections_stab}
\end{figure}

\begin{figure}
    \centering
    \includegraphics{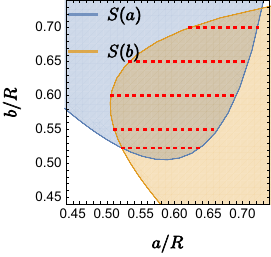}
    \caption{Same plot as in Fig. \ref{fig:two_intersections} (zoomed in).
    Horizontal dotted lines represent pairs $(a,b)$ associated with CO solutions that feature dynamical
    instability (see section \ref{sec:stability}). Horizontal lines are drawn at: $b/R =0.70$, $0.65$, $0.60$, $0.55$ and $0.52$.}
    \label{fig:two_intersections_stab_full}
\end{figure}

The stability character of the CO solutions can be represented in the rotational-state diagram $\cal D$. Figs. \ref{fig:four_intersections_stab} and \ref{fig:two_intersections_stab_full}
illustrate exemplary pairs $(a,b)$ belonging to a stable (horizontal continuous line) or unstable (dotted line) interval. In the two figures $a$ varies continuously along the intervals corresponding to a given value of $b$. 
Further numerical simulations suggest that, for $\Omega < \Omega_c$, the domain $\cal D$ does not exhibit any stability regions (see Fig. \ref{fig:two_intersections_stab_full}), thereby restricting the presence of stable CO solutions to the regime $\Omega \ge \Omega_c$. The complex form of the Jacobian matrix in Eq. \eqref{eq:dyn_system} and of the relevant eigenvalues does not allow to validate this result via an analytic approach.

\begin{figure}
    \centering
    \includegraphics[width=0.3\textwidth]{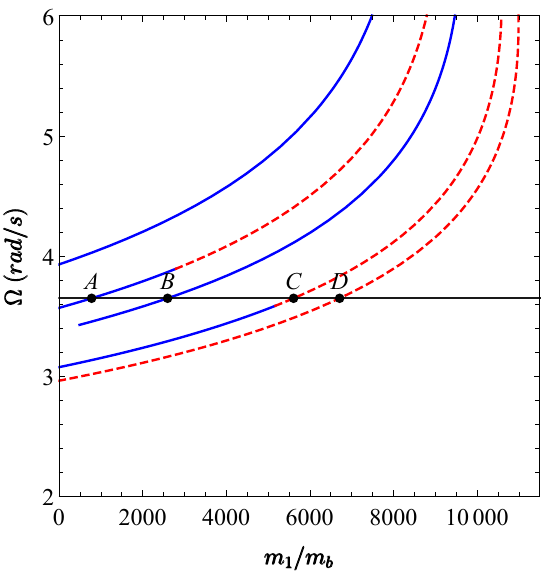}
    \caption{Same plot as in Fig. \ref{fig:omega_Branches_vs_m1}, (with a different aspect ratio). Pairs ($\Omega$, $m_1$) that feature dynamical stability (instability) are highlighted as continuous blue (dotted red) lines. Different curves corresponding to (from left to right) $a/R=0.38$, $0.78$, $0.45$, $0.52$, $0.65$, with $b/R=0.41$ and $N_A=10^5$. The curve for $a/R=0.45$ is interrupted at low $m_1$ values, where $\Omega$ does not satisfy Eq. \eqref{eq:paramEq2}.
    The horizontal line is drawn at $\Omega\simeq3.65$ $rad/s$, corresponding to the line at $b/R=0.41$ in Fig. \ref{fig:four_intersections_stab}.
    ``Large" values of $m_1$ are not physical since they lie outside the hypotheses of the point-like approximation.}
    \label{fig:omega_Branches_vs_m1_stab}
\end{figure}

Fig. \ref{fig:omega_Branches_vs_m1_stab} describes stable vs unstable domains of CO solutions by utilizing the same representation as in Fig. \ref{fig:omega_branches_stab},
where the continuous blue (red dashed) lines are associated with pairs $(\Omega,m_1)$ that ensure dynamical stability (instability). In the figure it is evident how the mass functions \eqref{eq:mass1} and \eqref{eq:mass2} are not monotonic in $a$ (or $b$). As a result, the stable, unstable and hybrid curves visible in Fig. \ref{fig:four_intersections_stab} form a sequence exhibiting an intermittent character when the parameter $a$ 
is varied. 
In Fig. \ref{fig:omega_Branches_vs_m1_stab}, for example, the horizontal line at $\Omega\simeq3.65$ $rad/s$, can be compared with the horizontal line at $b=0.41\;R$ in Fig \ref{fig:four_intersections_stab}. 
In the former, A, B, C, D are associated with increasing values of $m_1$, whereas, in the latter, the increasing values of $a$ are associated with the non-increasing mass-sequence B, C, D, A.

\begin{figure}
\centering
\includegraphics[width=0.3\textwidth]{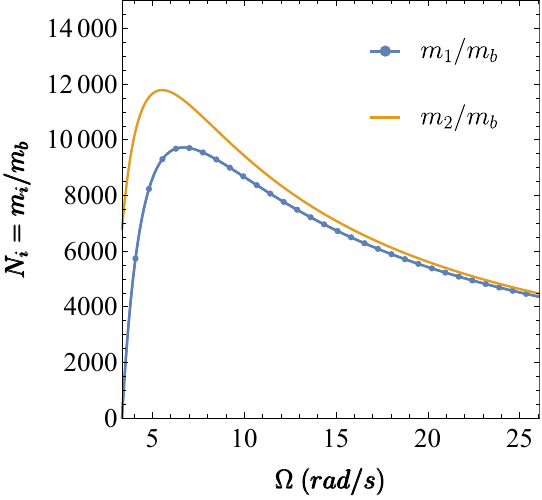}
\caption{\label{fig:masses_vs_omega}
Vortex masses $m_1$, $m_2$ as a function of $\Omega$ for $a/R$ $=0.70$, $b/R=0.30$, and $N_A=10^5$. The lowest frequency value $\Omega_{1m}\simeq3.35$ $rad/s$ corresponds to a single-mass VP with $m_1 =0$. For such solution $(a,b)$ belongs to the lower part of the $\cal D$ boundary.}
\end{figure}

Fig. \ref{fig:masses_vs_omega} describes the {\it mass imbalance} $m_2-m_1$ of the two vortices at a given point $(a,b) \in \cal D$, when $\Omega$ is increased. The lowest value of the frequency $\Omega =3.35\; rad/s$
corresponds to $(a,b)$ on the lower boundary of $\cal D$ where $m_1 (a,b)$ $\equiv 0$ and $m_2 (a,b)>0$. By increasing $\Omega$, the distance between the two surfaces $m_2(a,b)-m_1(a,b)$ decreases even if both $m_2(a,b)$ and $m_1(a,b)$ feature a (local) maximum. Both masses (and thus the imbalance) tend to zero for large $\Omega$. The parameter interplay characterizing a given CO solution shows how its geometry (orbit radii) is not affected by the change of $\Omega$ if the latter is compensated by a suitable variation of $m_2-m_1$. The fact that, in Fig. \ref{fig:masses_vs_omega}, $a>b$ and $m_1 <m_2$ for any $\Omega$, is a matter of chance depending on the values of $a$ and $b$.
In general, no elementary analytic relation can be found, \textit{a priori}, between the sign of $m_2-m_1$ and $a<b$ or $a>b$, two configurations potentially detectable at the experimental level. Nevertheless, for $\Omega > \Omega_c$, in view of the condition $m_1=m_2$ giving the transverse and diagonal branches of Fig.
\ref{fig:four_intersections_cut}, we can identify the two vertical (horizontal) counterposed ``triangles" where $m_2-m_1 <0$ ($m_2-m_1 >0$).  For $\Omega < \Omega_c$, the simplified geometry of $\cal D$ (see Fig.
\ref{fig:two_intersections}) entails that $m_2-m_1 <0$ ($m_2-m_1 >0$) for $a<b$ ($b<a$).

\section{Simulation of Gross-Pitaevskii equation}
\label{sec:GP_results}

In the following, we consider a mixture of $^{23}\mathrm{Na}$ (component $A$) and $^{39}\mathrm{K}$ (component $B$) in a trap of radius $R=50\,\mu m$. In the GPE simulations, we take the coupling constants to be 
$$
g_A=\frac{4\pi\hbar^2 a_A}{m_A}, 
\quad
g_B=\frac{4\pi\hbar^2 a_B}{m_B}, 
\quad
g_{AB}=\frac{2\pi\hbar^2 a_{AB}}{m_r},
$$
with $a_A\simeq 52.0\, a_0$ and $a_B\simeq 7.6\, a_0$ the intra-species s-wave scattering lengths, and $a_{AB}$ the scattering length between an $a$- and a $b$-atom. $a_0$ is the Bohr radius and the reduced mass $m_r$ obeys $1/m_r=1/m_A+1/m_B$. 
The harmonic oscillator characteristic length in the z-direction is $L_z = 2\times10^{-6}$. For the 2D modeling, the coupling parameters $g_i$ are normalized by $L_z$, and all the densities are planar.

\subsection{Code description}
\label{sec:code}

We prepare the initial state via the imaginary time evolution method.
Here, we rotate both the fluids at the selected $\Omega$.
Furthermore, we employ a rigid wall potential, crossing vertically the trap at $x=0$, to decouple the two vortices. This is switched off at the beginning of the real time simulation.
We stop the imaginary time right after the vortices have nucleated, to avoid as much as possible the redistribution of the $b$ masses. From this final state, the real time simulation starts.

\begin{figure}
\includegraphics[width=0.35\textwidth]{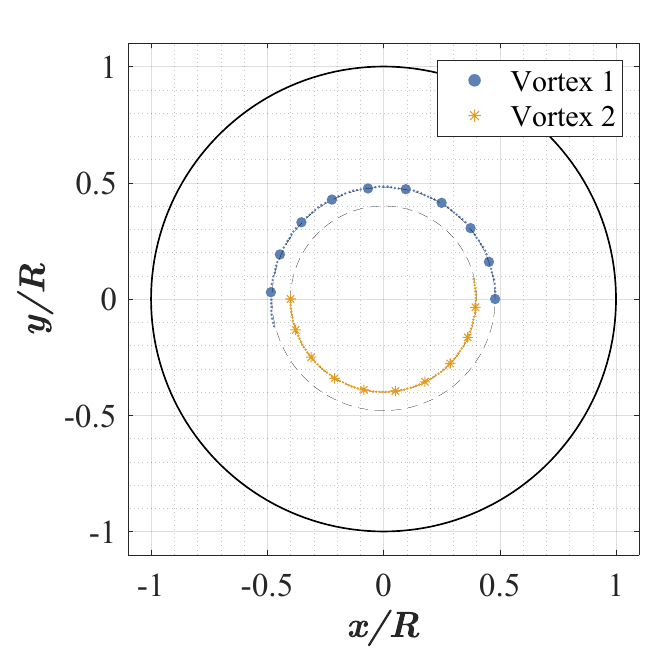}
\caption{GPE simulation of two imbalanced vortices. $N_A=10^5$, $a/R=0.48$, $b/R=0.4$, $N_1 \simeq 900$, $N_2\simeq 200$, $\Omega=3.36$ $rad/s$, $g_{AB}=2\sqrt{g_Ag_B}$. Run time: $1$ $s$, where the vortices rotate counter-clockwise. The dashed lines are the PL predictions, for some longer time.}
\label{fig:gpe_neat_ex}
\end{figure}

To extract the position of the vortices from our numerical simulations, we use an image processing tool that tracks the number-density minima in $n_A=|\psi_A|^2$.
An example of GPE trajectory is illustrated in Fig. \ref{fig:gpe_neat_ex}.

\subsection{GPE simulation}

\begin{figure}
         \centering
         \includegraphics[width=0.5\textwidth]{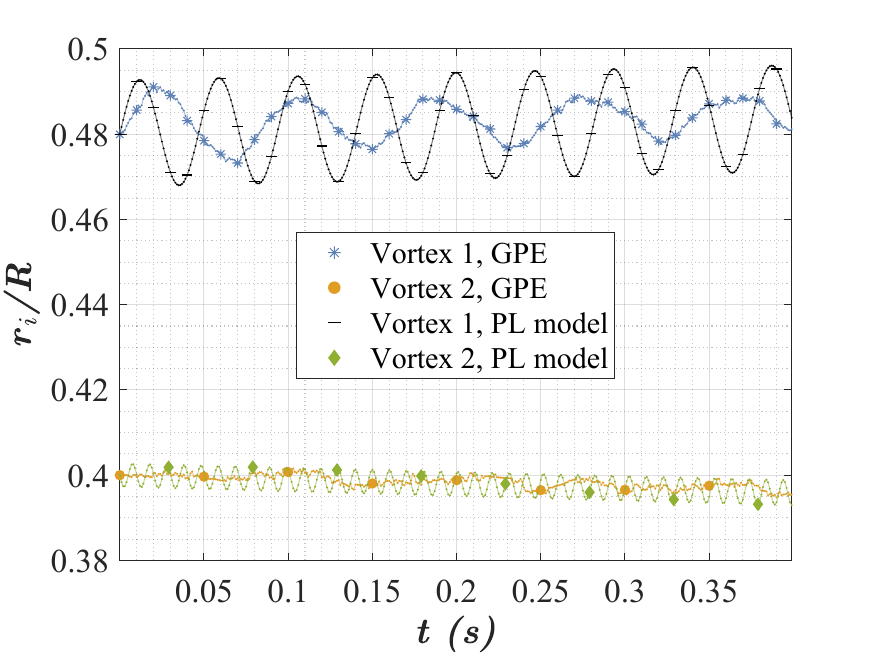}
         \caption{Comparison of the PL model prediction vs the GPE data for the vortex radial position, in the small-oscillations regime. Our purpose is to show the existence of the CO solutions with the GPE and the characterization of the small-oscillations around them. In this sense, the point-like model is reliable. $N_A=10^5$, $a/R=0.48$, $b/R=0.4$, $N_1\simeq 900$, $N_2\simeq 200$, $\Omega\simeq 3.36$ $rad/s$. $g_{AB}=2\sqrt{g_Ag_B}$. }
         \label{fig:comparison_PL_GPE_2}
\end{figure}

The agreement between GPE results and the predictions of our PL model for these simple orbits is good, as demonstrated by Fig. \ref{fig:gpe_neat_ex}.
Hence, the relations among the parameters of our two-vortex system found in section
\ref{sec:circular_solutions} are valid.
We then compare a PL solution in the small-oscillations regime with the GPE result. By means of the imaginary time evolution, we prepare our initial system, with the two imbalanced vortices. 
We compare our GPE result with the point-like model prediction. We find a quite good agreement as regards the stability features (see an example in Fig. \ref{fig:comparison_PL_GPE_2}), demonstrating that the point-like model is reliable. 
The quantitative disagreement of the model and the GPE trajectory is due the non-point-like character of the GPE system. The consistency of the GPE with the PL solutions improves when switching to regimes where the vortices are more point-like. This amounts to increasing $g_{AB}$ in parallel with $g_A$, so that the vortex profile is closer to a narrow Gaussian (see \cite{Bellettini2023}).
One could also resolve the disagreement by employing more accurate ansatzes as a starting point for the derivation of the PL model via the variational approach, at the price of ending up with more complicated expressions. Alternatively, by tuning effectively some parameters in the PL model, for a given system, one can obtain a very good agreement with the GPE trajectories.

Nonetheless, we are happy with the general qualitative and the quantitative agreement in the ``more point-like" regimes of the PL model with the GPE. This is in fact a dramatically simplified, classical model for some field non-linear equation. It serves here the purpose of proving the existence of some type of solutions of the GPE in the case of more vortices, and 
it correctly captures the dependency of the small-oscillations on the system parameters, such as the vortex masses. Therefore, we do not pursue the search of a perfect quantitative agreement between GPE and PL model.

\begin{figure}
         \centering
         \includegraphics[width=0.3\textwidth]{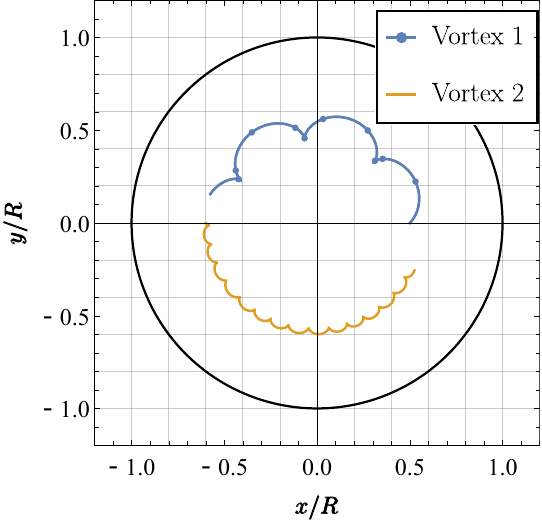}
         \caption{Point-like model prediction for the GPE solution in Fig. \ref{fig:comparison_large_oscillations_2}.  The GPE trajectory is pretty close to the PL model, even in the case of larger oscillations (the discrepancy is due to the non-point-like character of the massive-vortex profiles). Therefore, the type of solution we found, and their dependency on the physical parameters, is validated. $N_A=10^5$, $a/R=0.5$, $b/R=0.6$, $N_1\simeq 5200$, $N_2\simeq 1300$, $\Omega\simeq 2.77$. Run time: $1$ $s$.}
         \label{fig:comparison_large_oscillations_1}
\end{figure}

\begin{figure}
         \centering
         \includegraphics[width=0.35\textwidth]{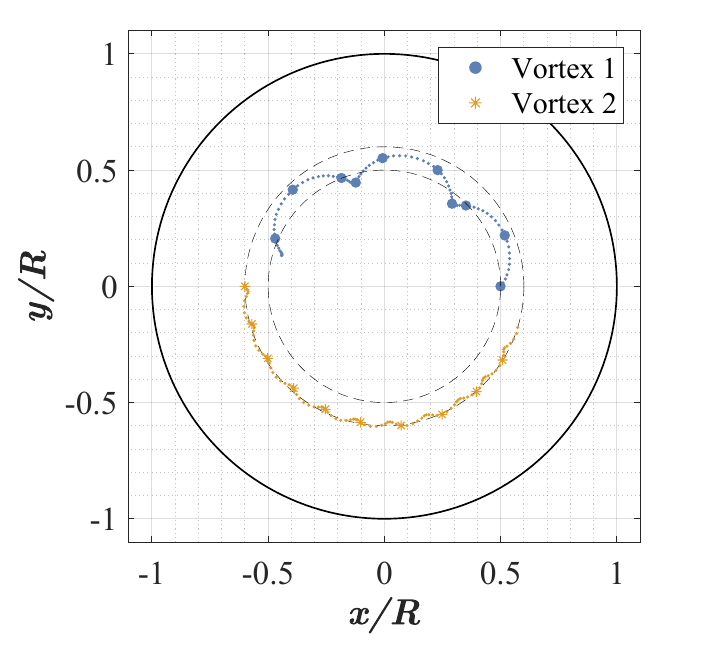}
         \caption{GPE solution relevant to the PL prediction in Fig. \ref{fig:comparison_large_oscillations_1} with $g_{AB}=3\sqrt{g_A g_B}$. }
         \label{fig:comparison_large_oscillations_2}
\end{figure}

We present an example of larger oscillations (see Figs. \ref{fig:comparison_large_oscillations_1} and \ref{fig:comparison_large_oscillations_2}).
The point-like model and the GPE predictions are quite similar, up to the radial oscillation frequency value of the first vortex. Hence, this is a validation of our PL model. It is visible how the more massive vortex manifests wider and slower radial oscillations in comparison with the second, lighter, one.

\section{Discussion and conclusions}
\label{sec:Conclusions}

We employ a point-like (PL) vortex model characterized by Lorenz-like equations to describe the dynamics of a pair of massive vortices in a binary mixture, confined in a 2D disk trap. 

In section \ref{sec:Model}, we find analytical solutions that two describe vortices with masses $m_1$ and $m_2$, moving along circular orbits exhibiting different radii $a$ and $b$, where the vortex positions have an angular shift of $\pi$. We discuss the dynamical stability conditions for the vortex trajectories in section \ref{sec:Normal_modes}, where CO solutions, featuring a uniform circular motion of vortices, have been shown to correspond to fixed points of the Hamilton equations in the rotating reference frame integral with the VP rotation.

In section \ref{sec:frequency_branches} we analyze the interplay of frequency $\Omega$,
the distinctive parameter of the rotational dynamics of CO vortex pairs, with the vortex masses, at given orbit radii $a$ and $b$.  

The two-branch plot of $\Omega$ as a function of $m_1$ or $m_2$ is analytically determined revealing i) the interval of the admitted frequency values, ii) the presence of upper limits for $m_1$ and $m_2$, and iii) the existence of CO solutions describing VPs where one of the two masses is zero. This type of solutions is found to correspond to the lower extreme of the $\Omega$ interval. This 
information is summarized in two numerical plots (associated to different choices of $a$ and $b$) showing $m_1$ ($m_2$) as a function of $m_2$ ($m_1$). 

In subsection \ref{sec:stability} we then explore the dynamical stability of the CO solutions, by exploiting plots where the precession frequency $\Omega$ depends on one of the two vortex masses and the relevant radius ($m_1$ and $a$ in the cases at hand). 
Fig. \ref{fig:omega_branches_stab} reveals a rich scenario of stable, unstable and hybrid plots, showing how the stability character of the CO solutions features a complex dependence from the interplay among masses and radii. This aspect is revisited from a more effective perspective in section  \ref{sec:mass_surfaces}.

In subsection \ref{sec:stability} we also
consider some properties of the small-oscillation motion, which is obtained by perturbing the CO solutions. Specifically,
the discussion on the eigenmodes of balanced VPs shows how trajectories characterized by small-oscillations, even when not providing a quantitative benchmark, serve as a signature of the vortex mass as much as the circular orbits themselves.
In this case, even with no information on the system parameters, we can infer the presence of a vortex mass from the appearance of relatively fast ``radial oscillations". In agreement with \cite{Bellettini2023}, the radial oscillations of a single vortex are found to increase in their amplitude, but decrease in their frequency, as the vortex mass increases. On the other hand, we find an oscillation mode that characterizes the VP also in the case of very small masses or even massless vortices. This occurs at longer time-scales with respect to the radial oscillations and provides a signature of small-mass motions.

The analysis developed in section \ref{sec:mass_surfaces} provides an exhaustive scenario of the rotational states associated with the CO solutions of VPs. The plots, at a given frequency $\Omega$,
of the vortex masses as functions of the radii $a$ and $b$ enables us to define the diagram $\cal D$ of the rotational states of VPs (determining, in turn, the existence domain of the CO solutions), and to highlight different classes of solutions. In addition to VPs with $a\ne b$ and different vortex masses, we find solutions describing 
single-mass VPs with $a\ne b$, and massless VPs both for $a\ne b$ and $a=b$. A further class, the VPs with $m_1=m_2$,
is shown to occur not only a for $a = b$ (diagonal branch), but also for $a\ne b$ (transverse branch).

The macroscopic changes in the structure of $\cal D$ when $\Omega$ is varied reveal a critical frequency $\Omega_c$ and the presence of two dynamical regimes $\Omega >\Omega_c$ and $\Omega <\Omega_c$.  Below $\Omega_c$, the significant changes of the  $\cal D$ boundary trigger the disappearance of the transverse branch of equal-mass solutions, shown in Fig. \ref{fig:four_intersections_cut}. In parallel, the frequency $\Omega_c$ supplies some useful information about the dynamical stability. As shown by numerical simulations (an analytical approach seems to be out of reach), $\Omega_c$ separates a regime where all the CO solutions are unstable ($\Omega < \Omega_0$) from that where extended regions of $\cal D$ correspond to stable CO solutions ($\Omega > \Omega_0$).
An analytic approach, instead, allows us to identify the minimum frequency $\Omega_0$ below which no VP rotation is admitted within the CO solutions. For $\Omega < \Omega_0$ the regions where either $m_1\ge 0$ or $m_2\ge 0$ have no overlap, and thus no CO solutions exist. 

\begin{figure}
    \centering
    \includegraphics[width=0.3\textwidth]{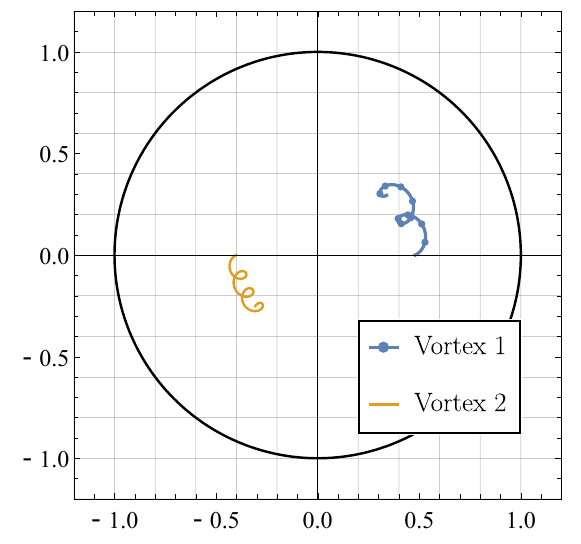}
    \caption{First oscillation mode of the imbalanced VP
    with $a/R=0.48$, $b/R=0.4$, $N_1=2000$, $N_2\simeq 1300$,
    $N_A=10^5$, $\Omega=3.46$ $rad/s$  (evolution time: $0.2$ s). The two vortices oscillate radially with no phase shift and at relatively high, different, frequencies. This type of oscillation is well visible for medium/large VP masses. The associated eigenvalue is $\omega_1$. 
    }
    \label{fig:radial_osc_out_phase}
\end{figure}

\begin{figure}
    \centering
    \includegraphics[width=0.3\textwidth]{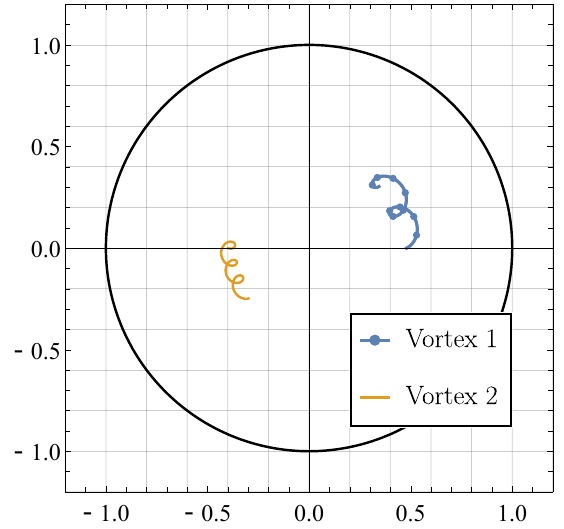}
    \caption{Second oscillation mode of the imbalanced VP, qualitatively similar to that of Fig. \ref{fig:radial_osc_out_phase}, except for the radial oscillations having a phase shift of $\pi$. The system data are the same.}
    \label{fig:radial_osc_in_phase}
\end{figure}

The analysis of section \ref{sec:mass_surfaces} highlights {\it i)} the deep link between the vortex masses and the adjustable geometry of the VP described by the orbit radii $a$, $b$, and {\it ii)} the link between the sign of $m_2-m_1$ and well defined subdomains of $\cal D$. More specifically, it suggests an indirect detection scheme for the vortex masses, given the orbit radii $a$ and $b$ and the rotation frequency. This
bypasses the difficulties of a direct measurement of the masses. Note that this scheme can be extended to fixing any other triplet within the five model parameters, to then determine the others. 

In section \ref{sec:GP_results}, we perform
simulations of the GP equations describing the binary condensate, to support the reliability of our CO solutions and of their dependency on the physical parameters.
For more complex trajectories, i.e. in the small-oscillations regime, some discrepancies of the PL model and the GPE arise due to the non-point-like character of the real system. Nevertheless, the general features of such trajectories 
(i.e. presence of a given oscillation mode, and dependency on the system parameters) are always at least qualitatively captured.

Concluding, the PL vortex model and the relevant CO solutions seem to provide a powerful tool for capturing the rotational dynamics of vortex-pair excitations with an adjustable geometry and depending on the vortex masses.
We expect that our results provide a useful guide for experimental applications where pairs of massive vortices are involved. This is the case not only for binary condensates, but also for systems including a thermal component \cite{Coddington2004, Griffin2009}, as well as in the presence of Andreev bound states in Fermi superfluids.
Also the case of vortex tracing active particles \cite{Giuriato2019, Griffin2020} falls into these applications.
Our future work will be focused on the study of more complex dynamical configurations and trapping geometries, and the possibility  to include dissipative terms in the PL mode which, as shown in reference \cite{Kwon2021}, are expected to better reproduce the vortex dynamics in a Fermi gas. 

\section*{Acknowledgements}
A. R. received funding from the European Union’s
Horizon research and innovation programme under the Marie Skłodowska-Curie grant agreement \textit{Vortexons} (no.~101062887). A. R. further acknowledges support by the Spanish Ministerio de Ciencia e Innovaci\'on (MCIN/AEI/10.13039/501100011033,
grant PID2020-113565GB-C21), and by the Generalitat
de Catalunya (grant 2021 SGR 01411).
Computational resources were provided by HPC@POLITO (http://hpc.polito.it).

\begin{appendix}

\section{Hamilton equations}
\label{sec:Hamilt}

Dynamical equations relevant to Hamiltonian \eqref{eq:Hamlab} are easily calculated by means of 
Poisson Brackets. One finds

\begin{equation}
    \dot{\bm{r}}_i=\frac{\bm{p}_i}{m_i}+\kappa_i\frac{m_A N_A}{2 m_i} \hat{e}_3\times \bm{r}_i
    \label{eq:Ham_eqs_ri}
\end{equation}

$$
        \dot{\bm{p}}_i  = \kappa_i\frac{m_A N_A}{2 m_i}\hat{e}_3\wedge\bm{p}_i - \kappa_i^2 \frac{m_A^2 N_A^2}{4 m_i}\bm{r}_i        
        $$
       $$             
       +\frac{m_A N_A}{4\pi}\Bigg[  2 \frac{\kappa_i^2 \bm{r}_i}{R^2-r_i^2}+
       \frac{R^2}{D(\bm{r}_i,\bm{r}_j)} 
       $$
       \begin{equation}
        \kappa_i \kappa_j \left(2
        \bm{r}_j -\frac{r_j^2}{R^2}\bm{r}_i
        +2\frac{(\bm{r}_i-\bm{r}_j)  
    D(\bm{r}_i,\bm{r}_j) }{R^2 (\bm{r}_i - \bm{r}_j)^2}\right)  \Bigg],
        \label{eq:Ham_eqs_pi}
\end{equation}
with $i\neq j$,
$D(\bm{r}_i,\bm{r}_j)= R^4-2R^2\bm{r}_i\cdot\bm{r}_j + {r_i^2 r_j^2}$.
Note that the case $\bm{r}_1=\bm{r}_2$ is in particular forbidden; this is not a problem since our ansatz assumes $\bm{r}_1$ and $\bm{r}_2$ always put off phase of $\pi$ radians.
\begin{figure}
    \centering
    \includegraphics[width=0.3\textwidth]{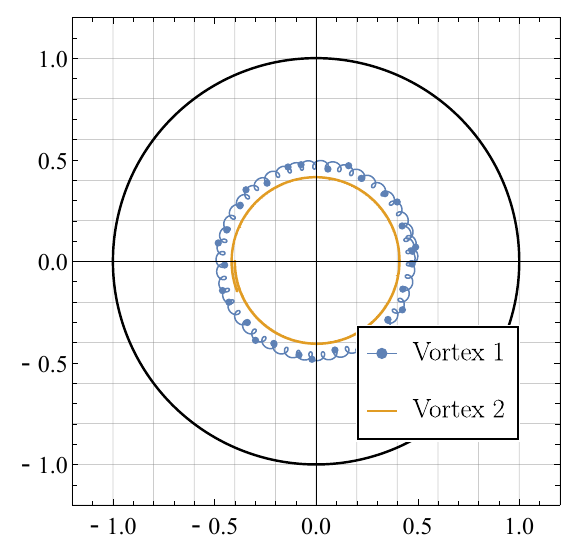}
    \caption{At longer times, the third oscillation mode of the VP also appears: the orbits radii slowly change so that the two trajectories do not wrap perfectly onto themselves after a revolution. This eigenmode is visible at small or zero vortex masses as well. $N_A=10^5$, $a/R=0.48$, $b/R=0.4$, $N_1=800$, $N_2=68$, $\Omega=3.35$ $rad/s$ (evolution time: $2$ s).}
    \label{fig:slower_osc}
\end{figure}

\begin{figure}
    \centering
    \includegraphics[width=0.5\textwidth]{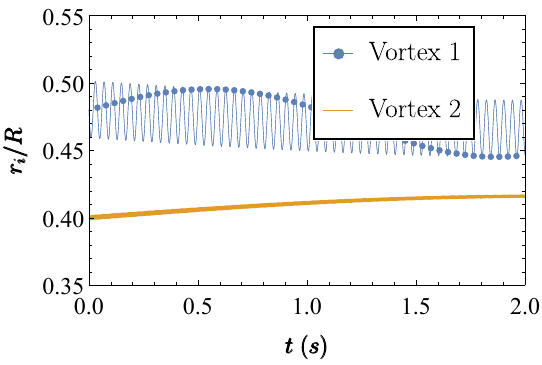}
    \caption{Radii $r_i$ of the vortex trajectories over a large time period (same data of Fig. \ref{fig:slower_osc}). The first vortex, with larger mass, oscillates radially at a detectable time-scale, while the less massive vortex $2$ features such fast radial oscillations, that they are hardly visible. The two radii also oscillate at a larger time-scale $1/\omega_3$, characteristic of the third oscillation mode. 
    }
    \label{fig:slower_osc_radii}
\end{figure}

\section{Normal modes analysis}
\label{appendix:small_oscillations}

For the study of the small-oscillations with respect to an equilibrium point, one considers small perturbations and the corresponding linearized system. We consider a reference frame rotating at frequency $\Omega$.
After linearizing with respect to $\bm{\xi}$, the Hamilton equations for the perturbed variables $\bm{z} = \bar{\bm{z}} +\bm{\xi} $ ($\bar{\bm{z}}$ is the vector describing the fixed-point coordinates), one obtains the linear system of equations for the perturbation $\bm{\xi}$
$$ 
\dot{\bm{\xi}} = \mathbb{J} \bm{\xi}, 
$$
where the Jacobian is $$
\mathbb{J}=\mathbb{E}\mathbb{H}.
$$ 
$\mathbb{H}$ is the (symmetric) Hessian matrix relevant to the Hamiltonian in the rotating reference frame
$$ 
\mathcal{H}_{rot}(\bm{\bar z}+\bm{\xi})=\mathcal{H}_{rot}(\bm{\bar z})
+
{\bf h }^T (\bm{\bar z}) \cdot \bm{\xi} 
+ \mathcal{H}_{rot, 2}
+O(\bm{ \xi }^2) ,
$$
where the second-order contribution to the rotational Hamiltonian reads
$\mathcal{H}_{rot,2}=
\frac{1}{2}
\bm{\xi}^T {\mathbb{H}} \bm{\xi}$
and the matrix elements of ${\mathbb{H}}$ are defined in Supplemental Material.
As expected, only the latter contributes to the equations of motion, since the 0-th order term is constant and the 1-st order contribution
depends on the vector
${\bf h }(\bm{\bar z})$ whose eight components correspond to the fixed-point Hamilton equations and thus are zero.
The linearized system above corresponds to $\dot{\bm{\xi}}
=
\{\bm{\xi}, \mathcal{H}_{rot,2}\}
=
\mathbb{E}\nabla_{\bm{\xi}} \mathcal{H}_{rot,2}$.
The elements of $\mathbb{H}$ (i.e. the second derivatives) divided by $2$ are the coefficients of the quadratic form  $\mathcal{H}_{rot,2}$.
If the eigenvalues of the Jacobian computed into a given point $(\bm{r}_1, \bm{r}_2)$ are pure imaginary, the fixed point is stable. 
Altogether, the eigenvalues of our two-vortex dynamical systems are eight.
Due to the nature of the Hamiltonian system, they are always paired as $\pm \sigma$.
Furthermore, due to the rotational invariance of our system, the third component of the angular momentum $L_3$ is conserved; thus, two eigenvalues are always zero. Again, since $\mathbb{J}$ is rotation-invariant, what matters for the stability analysis of the fixed point are the moduli of $(\bm{r}_1, \bm{r}_2)$.

\section{Eigenmodes}
\label{sec:appendix_eigenmodes}

The dynamical system associated with the VP features three characterisic oscillation modes. The first one, represented in Fig. \ref{fig:radial_osc_out_phase}, is triggered for example by symmetric perturbations of the vortex velocities. Here, the trajectory is characterized by radial oscillations of the two vortices, at relatively shorter time-scales. At $t=0$ both the radial coordinates of the vortices increase. In general, the two vortices exhibit different oscillation frequencies and amplitudes, which are the same in the balanced case $a=b$ and $m_1=m_2$. 
These kind of oscillations are the signature of non-negligible vortex masses and become slower and larger when increasing the vortex masses. Hence, are virtually invisible in the case of a small-mass VP.

Similar to the first mode is the second oscillation mode, represented in Fig. \ref{fig:radial_osc_in_phase}; this exhibits vortices with radial oscillations triggered, for instance, by an identical perturbation on the vortex velocities. At $t=0$ vortex-$1$ radial coordinate increases, while that of vortex $2$ decreases. Lastly, the slowest type of small-oscillations is illustrated in Figs. \ref{fig:slower_osc} and \ref{fig:slower_osc_radii} (period around $2\pi$).
They are  also visible in the case of massless or small mass vortices, and consist in a slow variation of the orbit radius. They appear at much larger time scales than the first two types. Here, the signature of a small vortex mass can be searched in the dependency of the frequency $\omega_3$ on $m_i$. This frequency increases at higher masses.

\end{appendix}

\end{document}